\documentclass[manuscript,screen,sigconf]{acmart}
\usepackage{soul}
\usepackage{enumitem}
\usepackage[most]{tcolorbox}
\usepackage{fontawesome}
\usepackage{wrapfig}
\usepackage{tabularx}
\usepackage[table]{xcolor}
\usepackage{makecell}
\usepackage{array}

\newcolumntype{P}[1]{>{\raggedright\arraybackslash}p{#1}}

\newcolumntype{M}[1]{>{\centering\arraybackslash\let\newline\\\arraybackslash\hspace{0pt}}m{#1}}

\usepackage{framed}
\usepackage[strict]{changepage}

\definecolor{formalshade}{rgb}{0.95,0.95,1}

\newenvironment{formal}{%
\begin{samepage}% <-- start samepage here
  \MakeFramed{\advance\hsize-\width\FrameRestore}%
  \noindent\hspace{-4.55pt}% disable indenting first paragraph
  \begin{adjustwidth}{}{7pt}%
  \vspace{2pt}\vspace{2pt}%
}
{%
  \vspace{2pt}\end{adjustwidth}\endMakeFramed%
  \end{samepage}% <-- end samepage here
}
\AtBeginDocument{%
  }

\setcopyright{acmlicensed}
\copyrightyear{2026}
\acmYear{2026}
\acmDOI{XXXXXXX.XXXXXXX}
\acmConference[EASE Companion 2026]{The 30th International Conference on Evaluation and Assessment in Software Engineering}{9–12 June, 2026}{Glasgow, Scotland, United Kingdom}
\begin{document}

%\title[Security Concerns in Generative AI Coding Assistants]{Security Concerns in Generative AI Coding Assistants: Insights from Stack Overflow, Reddit, and Hacker News}

\title[Security Concerns in Generative AI Coding Assistants]{Security Concerns in Generative AI Coding Assistants: Insights from Online Discussions on GitHub Copilot}

\author{Nicol\'{a}s E. D\'{i}az Ferreyra}
\authornote{Corresponding author}
\email{nicolas.diaz-ferreyra@tuhh.de}
\orcid{0000-0001-6304-771X}
\affiliation{%
  \institution{Hamburg University of Technology}
  \city{Hamburg}
  \country{Germany}
}

\author{Monika Swetha Gurupathi}
\email{monika.gurupathi@tuhh.de}
\affiliation{%
  \institution{Hamburg University of Technology}
  \city{Hamburg}
  \country{Germany}
}

\author{Zadia Codabux}
\email{zadiacodabux@ieee.org}
\orcid{0000-0001-6715-3341}
\affiliation{%
  \institution{University of Saskatchewan}
  \city{Saskatoon}
  \country{Canada}
}

\author{Nalin Arachchilage}
\email{nalin.arachchilage@rmit.edu.au}
\orcid{0000-0002-0059-0376}
\affiliation{%
  \institution{RMIT University}
  \city{Melbourne}
  \country{Australia}
}

\author{Riccardo Scandariato}
\email{riccardo.scandariato@tuhh.de}
\orcid{0000-0003-3591-7671}
\affiliation{%
  \institution{Hamburg University of Technology}
  \city{Hamburg}
  \country{Germany}
}

\renewcommand{\shortauthors}{D\'{i}az Ferreyra et al.}

\begin{abstract}

Generative Artificial Intelligence (GenAI) has become a central component of many development tools (e.g., GitHub Copilot) that support software practitioners across multiple programming tasks, including code completion, documentation, and bug detection. However, current research has identified significant limitations and open issues in GenAI, including reliability, non-determinism, bias, and copyright infringement. While prior work has primarily focused on assessing the technical performance of these technologies for code generation, less attention has been paid to emerging concerns of software developers, particularly in the security realm. \textbf{Objective}: This work explores security concerns regarding the use of GenAI-based coding assistants by analyzing challenges voiced by developers and software enthusiasts in public online forums. \textbf{Method}: We retrieved posts, comments, and discussion threads addressing security issues in GitHub Copilot from three popular platforms, namely Stack Overflow, Reddit, and Hacker News. These discussions were clustered using BERTopic and then synthesized using thematic analysis to identify distinct categories of security concerns. \textbf{Results}: Four major concern areas were identified, including potential data leakage, code licensing, adversarial attacks (e.g., prompt injection), and insecure code suggestions, underscoring critical reflections on the limitations and trade-offs of GenAI in software engineering. \textbf{Implications}: Our findings contribute to a broader understanding of how developers perceive and engage with GenAI-based coding assistants, while highlighting key areas for improving their built-in security features.

\end{abstract}

\begin{CCSXML}
<ccs2012>
   <concept>
       <concept_id>10002978.10003022.10003023</concept_id>
       <concept_desc>Security and privacy~Software security engineering</concept_desc>
       <concept_significance>500</concept_significance>
       </concept>
   <concept>
       <concept_id>10002978.10003029</concept_id>
       <concept_desc>Security and privacy~Human and societal aspects of security and privacy</concept_desc>
       <concept_significance>500</concept_significance>
       </concept>
   <concept>
       <concept_id>10002978.10003029.10011703</concept_id>
       <concept_desc>Security and privacy~Usability in security and privacy</concept_desc>
       <concept_significance>500</concept_significance>
       </concept>
   <concept>
       <concept_id>10011007.10011074.10011092.10011782</concept_id>
       <concept_desc>Software and its engineering~Automatic programming</concept_desc>
       <concept_significance>500</concept_significance>
       </concept>
 </ccs2012>
\end{CCSXML}

\ccsdesc[500]{Security and privacy~Software security engineering}
\ccsdesc[500]{Security and privacy~Human and societal aspects of security and privacy}
\ccsdesc[500]{Security and privacy~Usability in security and privacy}
\ccsdesc[500]{Software and its engineering~Automatic programming}

\keywords{Software Security, Security Concerns, Coding Assistants, Generative AI, Large Language Models, Topic Modeling}

\maketitle

\section{INTRODUCTION}
\label{sec:Introduction}

The application of Generative Artificial Intelligence (GenAI) technologies to software engineering has led to the emergence of innovative tools that are reshaping traditional programming workflows \cite{banh2025copiloting}. One of the most notable examples is GitHub Copilot\footnote{\url{https://github.com/features/copilot}}, a GenAI-driven code generation tool that delivers real-time, context-sensitive code suggestions in modern software development environments. Built on advanced Large Language Models (LLMs) like GPT-4o and GPT-4.1, Copilot supports the automated generation of entire functions, inline documentation, and test cases with minimal user input \cite{ebert2023generative}. By the end of 2024, this technology had been adopted by more than 50,000 organizations, with reported usage rates among developers exceeding 80\% in some cases, underscoring its growing impact on software engineering practices~\cite{gao2024quantifying}.

\subsubsection*{\textbf{Motivation}} Despite the benefits, prior work has raised concerns about the use of GenAI technologies to support software engineering tasks. Overall, several studies have identified significant quality issues in code generated by LLMs, especially regarding the prevalence of security weaknesses \cite{cotroneo2024vul,cotroneo2025devaic,zhi2025evaluating,tony2025tosem,fu2025security}. For instance, a recent study by \citet{fu2025security} found that approximately 30\% of Copilot's code suggestions in Python and JavaScript contained security flaws spanning multiple Common Weakness Enumeration (CWE) categories, including cross-site scripting and improper input validation. At their core, these issues stem from the data used to train and deploy GenAI technologies. That is, data scraped from publicly accessible code repositories containing security weaknesses, code anti-patterns, and other bad coding practices that LLMs reproduce verbatim \cite{hamer2024just,cotroneo2025devaic}. This not only threatens the overall quality of the software being developed but also raises important ethical and legal questions, including potential unauthorized use of copyrighted or sensitive code and liability for propagating errors and vulnerabilities \cite{stalnaker2024developer,carlini2022extracting}.

While previous investigations have primarily assessed the security standards of LLM-generated code, less attention has been paid to developers' perspectives on this. This includes identifying the security concerns, perceived risks, and challenges developers face when using GenAI-assisted code generation tools in their software engineering workflows. Platforms such as Reddit and Stack Overflow have proven valuable sources of insight into developers' mindsets, coding practices, and feedback on emerging technologies \cite{li2024unveiling}. Their relevance is reflected in numerous studies that leverage the content of these online communities (e.g., questions, answers, and discussion threads) to identify trends in the maintenance \cite{peruma2022refactor}, evolution \cite{zegers2025irresponsibility}, and security \cite{oishwee2024decoding} of information systems. Public attitudes and concerns in \textit{conversational} GenAI technologies (e.g., ChatGPT) have already been explored and analyzed through the lens of community-driven discussions \cite{xu2024public,ali2025understanding}.  However, to the best of our knowledge, limited research has leveraged these sources to uncover latent patterns of distrust, recurring security pain points, or to better understand how developers perceive the risks posed by GenAI-based coding tools. 

%... \hl{[HERE]} Platforms like Reddit, Stack Overflow, HackerNews and OpenAI’s community forums are rich sources of real-world input, where developers openly share their thoughts on Copilot’s behavior, report unexpected outputs, raise concerns, and sometimes caution others about potential privacy or security problems \cite{zegers2025accountability}. These discussions are \textbf{informal, firsthand, and grounded in experience}, making them highly valuable for understanding how practitioners actually view the \textbf{security risks} associated with GenAI-generated code.

%Yet, despite this wealth of content, no academic study has thoroughly mined and examined these conversations with a focus on identifying and categorizing \textbf{security-specific concerns}. While prior work \cite{zegers2025accountability} has applied topic modeling to developer discussions, the emphasis has typically been on broader topics like software accountability, not the specific security risks posed by GenAI code tools. Still, little efforts has been paid to uncover latent patterns of distrust, recurring security pain points, and to better understand how developers perceive the risks of GenAI-assisted coding.

\subsubsection*{\textbf{Contributions and Research Questions}}

This study explores security-related concerns expressed by software developers regarding the use of GenAI coding assistants. To this end, we curated and analyzed a dataset of security-focused discussions drawn from public online forums. Specifically, we collected discussions of GitHub Copilot from three popular Question-and-Answer (Q\&A) platforms and applied topic modeling in conjunction with qualitative analysis to identify recurring areas of concern. The research questions guiding this study are as follows:

\begin{itemize}[leftmargin=*]
    \item \textbf{RQ1:} \textit{What security, privacy, and trust concerns do developers express about GitHub Copilot in public technical forums?} To address this RQ, we curated a dataset of security-related posts and discussion threads about GitHub Copilot from Stack Overflow, Reddit, and Hacker News. We then applied BERTopic, a deep-learning-based topic modeling technique, in combination with a reflective thematic analysis to identify latent areas of concern across the dataset.
    \item \textbf{RQ2:} \textit{How do these concerns vary across developer communities such as Reddit, Stack Overflow, and Hacker News?} We examined the frequency and sentiment of each concern area across these communities to identify nuances in the way they are reported and addressed. We therefore aimed to assess whether platform norms and communication styles shape developers' security perceptions and their discussions of GenAI coding assistants.
\end{itemize}

This study resulted in a dataset of 383 Copilot-related discussions spanning four high-level concern areas, namely (i) \textit{Exposure and Integrity of Public Training Data}, (ii) \textit{Insecure Code Suggestions and Vulnerability Patterns}, (iii) \textit{Legal, Licensing, and Attribution Ambiguity}, and (iv) \textit{Developer Trust Erosion and Overdependence on GenAI}. To the best of our knowledge, \textbf{existing studies have not yet addressed security-related concerns about GenAI coding assistants across multiple Q\&A platforms}. Moreover, while prior developer-centered investigations have reported issues related to insecure code generation (e.g., \cite{klemmer2024using}), \textbf{two of the four identified concern areas have not been thoroughly examined in earlier work}. We present and discuss each area along with prospective research avenues to further enhance security, transparency, and accountability in the design of GenAI coding assistants.

%This study resulted in a dataset of 383 Copilot-related discussions spanning across four high-level concern areas, namely (i) \textit{Exposure and Integrity of Public Training Data}, (ii) \textit{Insecure Code Suggestions and Vulnerability Patterns}, (iii) \textit{Legal, Licensing, and Attribution Ambiguity}, and (iv) \textit{Developer Trust Erosion and Overdependence in GenAI}. We present and discuss each of them along with prospective research avenues to further enhance security, transparency, and accountability in the design of GenAI coding assistants.

The remainder of this paper is organized as follows. Section~\ref{sec:background} provides the background and summarizes the related work. We explain our research methodology in Section~\ref{sec:methodology}, followed by a detailed report of the study results in Section~\ref{sec:results}. Section~\ref{sec:discussion} reflects on the findings and provides implications for research and practice. We discuss the possible threats of our study and the mitigation strategies we adopted in Section~\ref{sec:limitations}. Section~\ref{sec:conclusion} concludes the paper and discusses future research directions.

\section{BACKGROUND AND RELATED WORK}\label{sec:background}

\subsubsection*{\textit{\textbf{Security Issues in LLM-Generated Code}}}

Since their emergence in the early 2020s, the quality of code generated by LLMs has been under close scrutiny \cite{fu2025security}. \citet{yetistiren2022assessing}, for instance, empirically evaluated the functional correctness, validity, and efficiency of Copilot-generated code using HumanEval \cite{chen2021codex}, a dataset of 164 Python programming tasks comprising function signatures, docstrings, and unit tests. Their results showed that Copilot could generate valid, executable code around 90\% of the time, while in 80\% of the cases it delivered either correct or partially correct implementations. \citet{pearce2022copilot} also leveraged and extended the HumanEval dataset to identify issues in security-critical coding tasks and found that 40\% of Copilot's solutions contained CWEs (i.e., security weaknesses) ranked among MITRE's Top-25 most dangerous ones. In the same vein, \citet{tony2025tosem} explored whether the use of different prompting techniques can influence the security levels of LLM-generated code using LLMSecEval \cite{tony2023msrtool}, a dataset of programming tasks expressed in natural language. They showed that, although techniques like Recursive Criticism and Improvement (RCI) and Chain of Thought (CoT) can reduce the number of CWEs across different LLMs (e.g., GPT-3.5, GPT-4, and LLama), they are still prone to introducing certain CWEs like \textit{code injections} (CWE-94) and \textit{hard-coded passwords} (CWE-259).% Furthermore, a similar behavior has been reported in a recent study by \citet{fu2025security} on Python and JavaScript code snippets generated by Copilot and two other LLMs (i.e., CodeWhisperer and Codeium).

Code and data memorization is another frequently reported issue that often raises concerns among software practitioners \cite{wu2025empirical}. Overall, it refers to LLMs' tendency to mirror or duplicate their training data, leading to the verbatim reproduction of copyrighted code snippets, proprietary algorithms, or vulnerable code patterns. \citet{niu2023codexleaks}, for example, demonstrated that carefully crafted prompts can induce Copilot to reveal sensitive information memorized from public GitHub repositories. Their study revealed that Copilot memorized personal data (e.g., email addresses or phone numbers) and even passwords embedded in publicly available database queries or JSON files during its training. Another study by \citet{yang2024unveiling} on CodeParrot — an open-source GPT-2-based LLM — revealed that the distribution of memorized content is closely linked to the type of prompts provided. In other words, the way developers phrase their requests can bias the model toward reproducing specific categories of training data, such as configuration files, class definitions, or import statements. Recent work highlights that LLMs can disclose training data even when not intentionally prompted to do so \cite{rabin2025malicious}. In turn, there is a non-zero chance that non-malicious users may encounter sensitive information in LLM outputs generated in response to a mundane programming task.

%highlighted/concluded that ... built a taxonomy of memorized content, prompts affect the distribution of memorized content, LLMs producing longer outputs have a higher tendency towards memorization

%these are on the quality of Copilot. Then mention the ones comparing LLMs and other things like prompt engineering.

%security issues

%memorization issues, hallucinations?

\subsubsection*{\textit{\textbf{Emerging Concerns in GenAI Technologies}}}
Prior work has examined public perceptions of GenAI technologies, revealing recurring concerns around privacy, security, and trust among end users. A study by \citet{koonchanok2024public} found that cybersecurity was among the most frequently discussed topics on the $\mathbb{X}$ platform (formerly Twitter) regarding ChatGPT. These discussions were predominantly negative, portraying ChatGPT as a disruptive technology that threatens privacy and raises security risks. Similar findings were reported by \citet{okey2023investigating}, who observed concerns among $\mathbb{X}$ users regarding the possibility of ChatGPT being exploited as a hacking tool (e.g., to generate malicious code) or for facilitating social engineering attacks. Concerns about privacy and data misuse have also been raised online, underscoring the importance of transparent data-handling practices to enhance the trustworthiness of GenAI systems \cite{al2024chatgpt}. Furthermore, a recent study of the ChatGPT Reddit community \cite{ali2025understanding} identified concerns across all stages of the data lifecycle (i.e., collection, use, and retention). Overall, users seem worried about their behavior being monitored and sensitive data (e.g., work-related information) being exposed to unauthorized parties in corporate settings.

Recent investigations have also begun to explore how software developers perceive and engage with GenAI technologies in their day-to-day coding activities. \citet{nguyen2025generative}, for instance, conducted a series of focus groups and identified several open challenges around the use of GenAI for software engineering. Among the key findings, the study emphasizes the need to develop new skills to enable practitioners to deliver robust LLM-based implementations, as well as the importance of human oversight to ensure trustworthy software solutions. In the same vein, interviews and exploratory case studies with development teams \cite{dolata2024development,WuGehbauer2024,banh2025copiloting} have revealed quality-savvy practices in the wild, such as providing context-relevant information to LLMs (e.g., using prompt engineering techniques) and closely monitoring their outputs through Static Application Security Testing (SAST) tools. Nevertheless, these practices remain challenging since the stochastic and opaque nature of LLMs makes it hard to establish common criteria for quality assessment \cite{dolata2024development}. Moreover, the possibility of technologies such as GitHub Copilot gaining access to a wide range of proprietary code and documentation creates significant barriers to the adoption of GenAI solutions across business contexts \cite{banh2025copiloting}. This and other security-related issues have also been reported in a recent study by \citet{klemmer2024using}, which used semi-structured interviews and analysis of Reddit data. Still, developer-centered insights remain limited in this regard and call for further investigation into the specific types of security concerns arising from GenAI-assisted coding practices. In particular, to the best of our knowledge, \textbf{prior work has not yet systematically characterized security concerns across multiple developer platforms nor provided fine-grained insights into their types and distribution}.

%Yet, developer-centered insights remain limited in this regard and call for further investigation into the specific types of security concerns arising from GenAI-assisted coding practices.

% to gain a deeper understanding of the security concerns around the use of GenAI coding assistants.

%Yet, current studies provide little insight into the security of GenAI coding assistants, underscoring the need for further developer-centered investigations in this area.

\section{METHODOLOGY}\label{sec:methodology}

\begin{figure}[t]
    \centering
    \includegraphics[width=\linewidth]{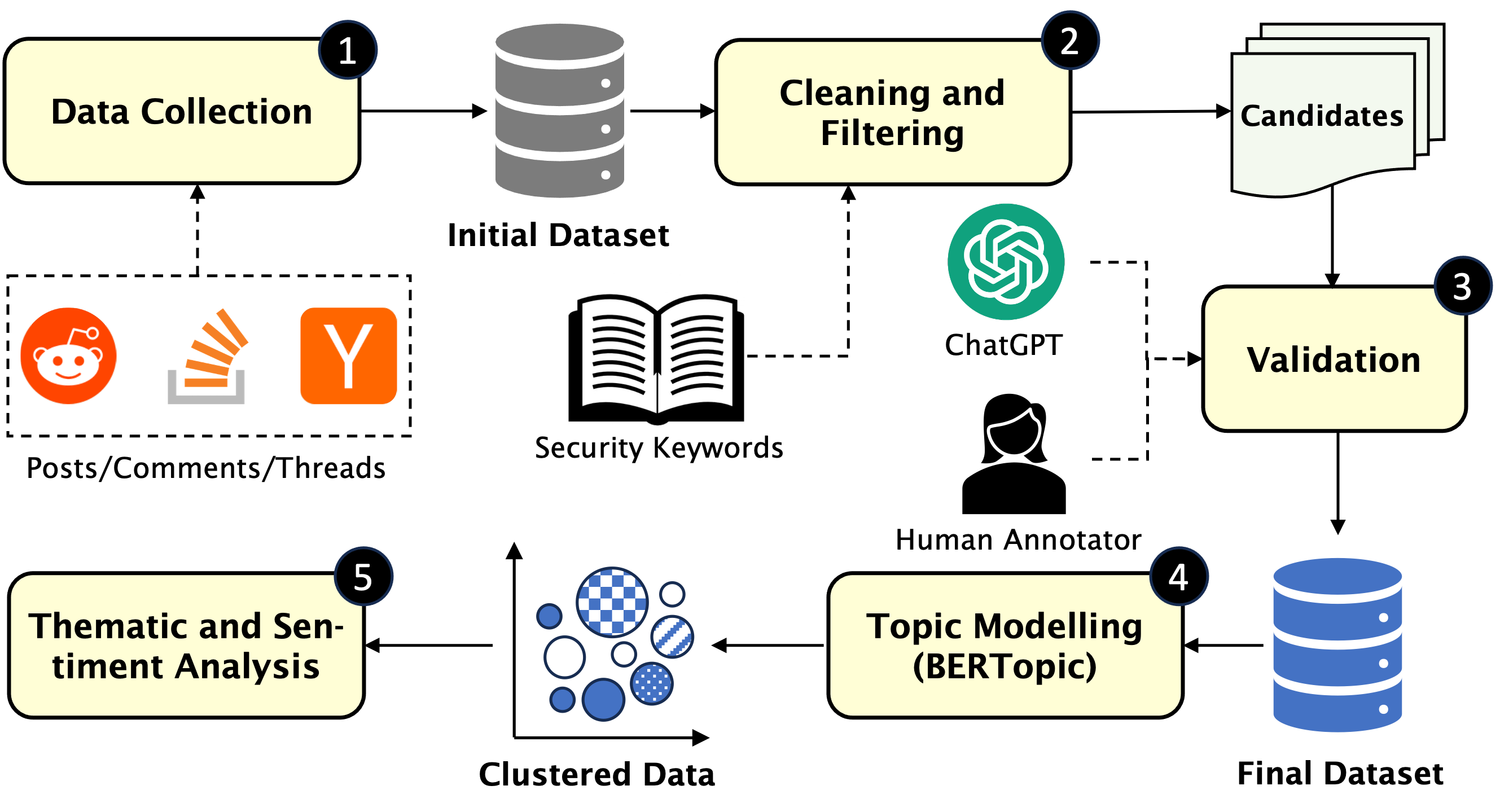}
    \caption{Study Design.}
    \label{fig:methodology}
\end{figure}

To answer the RQs introduced in Section~\ref{sec:Introduction}, we curated a dataset of online posts, comments, and discussion threads addressing security issues in GitHub Copilot from three public online forums: Stack Overflow, Reddit, and Hacker News. As shown in Fig.~\ref{fig:methodology}, we clustered these discussions using BERTopic, a state-of-the-art topic modeling technique, and identified salient concern categories through thematic analysis. In the following subsections, we describe the different steps of our study design along with the techniques employed during the different data acquisition and
processing activities.

\subsubsection*{\textbf{STEP 1: Data Collection}} We selected Stack Overflow, Reddit, and Hacker News as our primary sources of data as they are known for fostering active communities of users who engage in conversations about software development, coding tools, and best practices \cite{li2024unveiling, chavan2024analyzing, antelmi2023age}. Stack Overflow data (i.e., questions and their associated answers) was collected through the Stack Exchange API by querying posts tagged with \texttt{[github-copilot]}. This tag was chosen to ensure relevance to discussions focused specifically on GitHub Copilot, as it is consistently used by the Stack Overflow community to label content related to the tool. Relevant Hacker News discussions were retrieved using the official Hacker News Firebase API and identified by searching for explicit mentions of \texttt{``GitHub Copilot''}. All nested comment threads within the matched posts were recursively extracted to preserve conversational context. For Reddit, we mined relevant discussions from subreddits on software development (i.e., \texttt{r/programming}, \texttt{r/learnprogramming}, \texttt{r/opensource}, and \texttt{r/github}) and AI (i.e., \texttt{r/ArtificialIntelligence}) using the platform's API. As with Hacker News, these discussions were identified by searching for the \texttt{``GitHub Copilot''} keyword in the body of posts and comments within the selected subreddits. By the end of this step --conducted in May 2025-- we had obtained an \textit{initial dataset} of \textbf{14,253 Copilot-related comments} from which 2,227 were retrieved from Stack Overflow, 3,667 from Hacker News, and 8,349 from Reddit.

\subsubsection*{\textbf{STEP 2: Cleaning and Filtering}} We followed a keyword-based search strategy to detect security-relevant comments in our \textit{initial dataset}. For this, we used a well-established list of security terms from \citet{croft2022empirical} which comprises 266 security keywords (e.g., `phishing', `thread-safe'), making it one of the most extensive ones documented in the literature. Although this type of filtering has known limitations (e.g., it may yield false positives), we considered it an efficient first-pass filter at this stage, given the size of our dataset and the relatively short text length of the discussions on the investigated platforms. As shown in Table~\ref{tab:filter_data}, the keyword search returned 3,360 candidate posts (i.e., potentially security-relevant), which were then manually validated.

\begin{table}
    \centering
    \caption{Initial and Filtered Dataset Counts.}
    \begin{tabular}{l|r|r} \toprule
        \textbf{Source} & \# \textbf{Discussions} & \# \textbf{Security Candidates}\\ \hline
        Stack Overflow & 2,227 & 309\\ 
        Hacker News & 3,667 & 390\\ 
        Reddit & 8,349 & 2,659\\ \hline
        \textbf{TOTAL} & \textbf{14,253} & \textbf{3,360}\\ \bottomrule
    \end{tabular}
    \label{tab:filter_data}
\end{table}

\subsubsection*{\textbf{STEP 3: Validation of Security Candidates}} Following the initial keyword-based filtering, a size-dependent validation protocol was established to assess the relevance of candidate entries. We thereby adopted a dual strategy that scaled the depth of manual review to the dataset size, allowing exhaustive validation in smaller partitions and more selective filtering in larger ones.

\begin{itemize}[leftmargin=*]
    \item \textbf{Case 1 -- Exhaustive Manual Validation ($N \leq 400$)}:  
    For partitions of manageable size, a complete manual validation was performed to ensure maximum coverage and thematic accuracy. Specifically, Stack Overflow ($N = 304$) and Hacker News ($N = 390$) each contained fewer than 400 entries, making them feasible for exhaustive validation. Hence, every comment within these partitions was manually reviewed by an author to determine its relevance (i.e., whether it voiced security, privacy, or trust concerns related to GitHub Copilot).
    \item \textbf{Case 2 -- Keyword Refinement and Targeted Filtering ($N > 400$)}:  
    For the Reddit partition ($N = 2{,}659$), where full manual review was impractical, we adopted a two-stage keyword validation pipeline. First, keywords that consistently yielded true positive cases in the manually validated Stack Overflow and Hacker News partitions (Case 1) were retained and used to re-filter the Reddit partition. The resulting subset was then reviewed in full by an author to ensure its relevance and to discard false positives.
\end{itemize}

During validation, each candidate post was labeled either as \textit{relevant} (i.e., it clearly discusses security, privacy, or trust-related concerns) or \textit{irrelevant}. For exhaustive manual validation (Case~1), we used ChatGPT (v4.0) as a second annotator and compared its output to the author's labels. In cases of disagreement, the author manually revised their original annotation and modified it as deemed appropriate. The resulting Cohen's Kappa was 0.91 for Stack Overflow and 0.94 for Hacker News, indicating an almost perfect inter-rater agreement between the author and ChatGPT. Overall, six security terms (i.e., \texttt{violate}, \texttt{privacy}, \texttt{insecure}, \texttt{leak}, \texttt{trust}, and \texttt{security}) from the original list of 266 lead consistently to true positive cases during the validation of both partitions. The candidates obtained after re-filtering the Reddit subset with these keywords were double-checked by another author. \textbf{By the end of this step, we had obtained a validated dataset of security concerns containing 383 instances (20 from Stack Overflow, 170 from Hacker News, and 193 from Reddit) spanning June 2021 to March 2025}.

\subsubsection*{\textbf{STEP 4: Topic Modeling}} We analyzed the resulting dataset using BERTopic, a topic modeling technique that leverages transformer-based embeddings and density-based clustering to organize text segments into semantically related groups. This technique allowed us to arrange closely related security discussions into coherent clusters, enabling a structured exploration of developers' concerns surrounding GitHub Copilot. Although the final dataset comprises 383 individual comments, many of them are relatively long and embedded in broader conversational contexts (e.g., on Hacker News), resulting in substantial semantic depth per data point. Accordingly, we use BERTopic as an \textit{exploratory structuring mechanism} to organize semantically related comments prior to qualitative synthesis, rather than as a standalone method for deriving definitive topic structures.

Before applying topic modeling, we performed standard text preprocessing steps, including removing placeholder tokens, URLs, email addresses, and excessive whitespace, to reduce noise and improve the quality of the input data. Each data point was encoded into a dense, high-dimensional vector using Sentence-BERT (SBERT), followed by dimensionality reduction with Uniform Manifold Approximation and Projection (UMAP). We then applied HDBSCAN for density-based clustering with a minimum cluster size of 10 and used a CountVectorizer to extract representative keywords for each cluster. This process yielded 11 coherent clusters of security-related discussions.

\subsubsection*{\textbf{STEP 5: Thematic and Sentiment Analysis}}
The resulting clusters were further examined qualitatively through a \textit{reflective thematic analysis} \cite{cruzes2011recommended} to identify overarching categories of security concerns. To this end, we employed a lightweight open coding approach in which one author reviewed and assigned descriptive labels to the discussions within each cluster. They then inductively extracted emerging themes and recurring patterns to capture core concern areas. The resulting codes and themes were documented in a shared \textit{codebook} to ensure consistency and traceability throughout the analysis. 

Clusters containing only a small number of discussions ($N \leq 30$) were examined in full, whereas representative samples were taken from the larger ones. After open coding, some were merged due to substantial thematic overlap. By the end of this process, we had identified \textbf{four major categories of security concerns} surrounding the use of GitHub Copilot from a developer-centered perspective. As a final validation step, a second author with extensive experience in software security reviewed the identified concern areas and their associated clusters to assess coherence, internal consistency, and alignment with the underlying discussions, leading to minor refinements where appropriate.

To complement this qualitative assessment, we conducted sentiment analysis of all comments to examine how attitudes toward Copilot's security implications varied across platforms. Each comment was assigned a continuous polarity score between -1 (strongly negative) and +1 (strongly positive) using the transformer-based model \texttt{cardiffnlp/twitter-roberta-base-sentiment}, enabling comparison of sentiment distributions across Reddit, Stack Overflow, and Hacker News.

\subsubsection*{\textbf{Ethical Considerations}} Each stage of the research process, from data collection to annotation and analysis, was critically reviewed to minimize risk to individuals, uphold transparency, and align with ethical standards in software engineering research \cite{gold2022ethics}. All datasets used in this study were sourced from publicly accessible developer discussion forums: Stack Overflow, Hacker News, and Reddit. In accordance with common research practice \cite{codabux2024teaching,casari2023beyond}, only publicly visible data was collected, and no attempts were made to access private messages, deleted content, or user-specific histories. Furthermore, data mining procedures adhered to platform-specific usage policies to ensure compliance with the terms of service. Collected posts were used solely for research purposes and aggregated to examine broader trends in developer discourse regarding the use of GenAI coding assistant tools. To preserve user privacy, no Personally Identifiable Information (PII) was extracted or retained. Usernames, timestamps, comment IDs, and profile metadata were excluded from the final dataset. All analyses focused exclusively on the textual content of posts, ensuring that individuals could not be identified, re-identified, or profiled based on the content.

\begin{tcolorbox}[width=\linewidth,colback={gray!20},boxsep=0.5ex]    
All study materials, including the topic modeling Python scripts, the code book, and the annotated spreadsheets, are publicly available in the paper's \textbf{Replication Package\footnotemark}.
\end{tcolorbox}
\footnotetext{\url{https://doi.org/10.5281/zenodo.19484951}}

\section{RESULTS} \label{sec:results}

\begin{figure}[t]
    \centering
    \includegraphics[width=\linewidth]{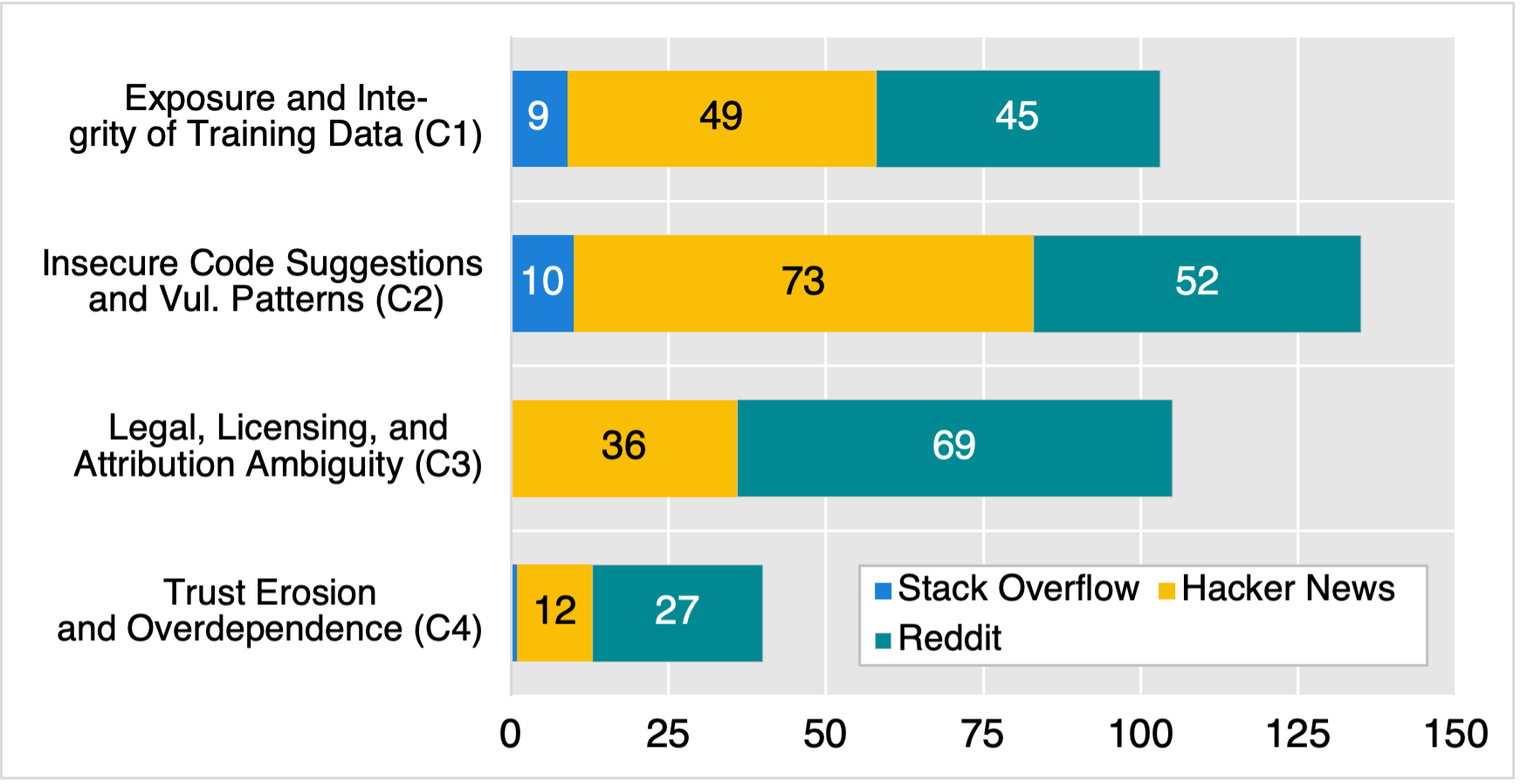}
    \caption{Provenance Distribution Across Clusters.}
    \label{fig:distribution}
\end{figure}

%As shown in Table~\ref{tab:theme_summary},
We identified four high-level areas that capture prominent Copilot-related security concerns discussed in our dataset, namely (i) \textit{``Exposure and Integrity of Public Training Data,''} (ii) \textit{``Insecure Code Suggestions and Vulnerability Patterns,''} (iii) \textit{``Developer Trust Erosion and Overdependence on GenAI,''} and (iv) \textit{``Legal, Licensing, and Attribution Ambiguity.''} The contribution of each platform to these areas is displayed in Fig.~\ref{fig:distribution}. As observed, users of both Hacker News and Reddit expressed concerns across all areas, whereas Stack Overflow users did not raise concerns about trust or over-reliance. In the following sections, we introduce each identified concern area in detail while illustrating key points with examples from our dataset. We further complement this analysis by highlighting the nuances in sentiment observed across the investigated platforms.

%Next, we elaborate on the five categories.

\subsection{Security Concern Areas (RQ1)}
% \textit{RQ1: What security, privacy, and trust concerns do developers express about GitHub Copilot in public technical forums?}
% \textcolor{blue}{5 themes namely:\\
% 4.4 Exposure of Secrets and Training Data Leakage \\
% 4.5 Insecure Code Suggestions and Vulnerability Patterns \\
% 4.6 Developer Trust Erosion and Overdependence on AI \\
% 4.7 Legal, Licensing, and Attribution Ambiguity \\
% 4.8 Prompt Injection and Training Data Poisoning}

% \textbf{Clustering.}
% First, we report the results of topic modeling and thematic clustering to better understand the structure and content of developer concerns related to GitHub Copilot. Our analysis yielded eleven distinct topic clusters (Topic 0 through Topic 10), each representing a semantically coherent group of comments. These clusters served as the foundation for subsequent open coding and thematic analysis.

% Then, the clusters served as the input for open coding, enabling the identification of granular concerns at the individual comment level. The data was structured into semantically coherent topic clusters using BERTopic and then refined through an open coding process. This 

\subsubsection*{\textbf{\textcircled{\small{1}} Exposure and Integrity of Public Training Data.}}
The most common concern is the potential for GitHub Copilot to expose sensitive or proprietary content memorized during its training phase. These discussions elaborate on a spectrum of security-related anxieties, ranging from unintended data memorization to the ethical implications of leaking proprietary logic or secrets. They reflect both technical issues, such as \textit{model inference attacks} (i.e., the ability to reconstruct training data through repeated or cleverly crafted queries), and broader normative challenges regarding transparency, consent, and accountability in GenAI-assisted development workflows. Users frequently debated the likelihood that Copilot would reproduce publicly available code verbatim and the consequences of incorporating such snippets into proprietary projects.

\begin{formal}\faPencil~
\textit{``...in case of enterprise code development, where code must remain strictly confidential, can GitHub Copilot save any sort of code (entirely or even just snippets) and make it public with suggestions?'' (Reddit user)}
\end{formal}

We also observed several references to risks in this area stemming from the manipulation of Copilot's behavior through data poisoning. That is, on the possibility that public repositories used for training could be intentionally seeded with insecure or malicious code snippets to influence Copilot's future outputs. Mentions of prompt injection attacks (i.e., deceiving Copilot with malicious instructions to bypass security guardrails) were also noted, though less frequently. Such exchanges suggest increasing awareness among developers about existing vulnerabilities in the mechanisms that govern how GenAI coding assistants learn and respond. Furthermore, they resonate with ongoing challenges in adversarial machine learning, acknowledging that public code repositories could serve as a potential attack vector to intentionally degrade Copilot's long-term performance and reliability.

\begin{formal}\faPencil~
\textit{``So...say Microsoft retrained Copilot on code-only explicitly marked as open-source. As an activist or vandal, you could start publishing proprietary code with fraudulent license files to pollute Copilot again. This could be terribly fun.'' (Hacker News user)}
\end{formal}
%\textcolor{blue}{Without stricter transparency, sanitization of training data, and real-time safeguards such as secret detection or prompt-based warnings, trust in Copilot’s deployment within security-critical environments is likely to remain constrained. This theme highlights the importance of secure data governance throughout the lifecycle of AI-assisted development tools.}

\subsubsection*{\textbf{\textcircled{\small{2}} Insecure Code Suggestions and Vulnerability Patterns.}}
A significant number of concerns focused on the quality of code generated by GitHub Copilot, specifically insecure defaults, insufficient security mechanisms, and vulnerability-prone design patterns. Developers attributed this to either insufficient sanitization of the training corpus or the over-representation of insecure examples in public repositories. These discussions range from specific vulnerability reports, such as SQL injection or broken authentication, to broader critiques of Copilot's apparent lack of security awareness. While Copilot has been widely praised for improving development speed, these findings suggest that its use may compromise application security, particularly when outputs are accepted without manual verification or hardening.

\begin{formal}\faPencil~
\textit{``I think that AI won't replace programmers, at least for now. All those codes written by AI still need review and more attention within security.'' (Hacker News user) }
\end{formal}

\begin{formal}\faPencil~
\textit{``...I've seen enough code on Github with obscure security flaws to be wary of any code it generates...As the model doesn't have any comprehension of the code itself, it's likely to suggest code because it's common rather than good.'' (Reddit user)}
\end{formal}

Although many view Copilot as a powerful tool to accelerate coding and streamline repetitive tasks, they frequently note that its suggestions often omit essential security features or rely on implementations that lack defensive programming practices. Such experiences raise doubts about the reliability of Copilot's recommendations in real-world, security-sensitive contexts. Across discussions, a shared sentiment emerges: GenAI coding assistants tend to prioritize functionality and convenience over security, leaving key safeguards unaddressed unless explicitly requested.

\begin{formal}\faPencil~
\textit{``Copilot was trained on code without any quality metric. So it will happily reproduce all bugs, security issues, and deprecated API usage found in any dead GitHub project.'' (Stack Overflow user)}
\end{formal}

\subsubsection*{\textbf{\textcircled{\small{3}} Legal, Licensing, and Attribution Ambiguity}} 
Another recurring topic in developer discussions concerns the legal status, licensing implications, and attribution challenges associated with code generated by GitHub Copilot. Particularly, about the provenance of such code and whether its reuse might introduce legal conflicts (e.g., copyright infringement or code plagiarism), especially within organizations that operate under strict audit and compliance policies. Unlike the technical concerns voiced in the two prior cases (i.e, ``Exposure and Integrity of Public Training data'' and ``Insecure Code Suggestions and Vulnerability Patterns''), these discussions reflect a structural opacity in the design of GenAI coding assistants: users are unable to verify where a given suggestion originates or whether its reuse carries legal or ethical obligations. 

\begin{formal}\faPencil~
\textit{``What happens when someone puts code up on GitHub with a license that says `This code may not be used for training a code generation model'? Is GitHub actually going to pay any attention to that, or are they just going to ingest the code and thus violate its license anyway? If they go ahead and violate the code's license, what are the legal repercussions for the resulting model?...'' (Hacker News user)}
\end{formal}

\begin{formal}\faPencil~
\textit{``In my opinion, it violates most licenses... Even licenses like MIT require to give attribution, which Copilot isn't doing. The GPL requires that you license under GPL if you include any part of the code...'' (Reddit user)}
\end{formal}

Such a black-box behavior, combined with the lack of code authorship metadata, makes it hard to determine whether its reuse may (or may not) conflict with project-specific and organizational policies. Several users also expressed unease about not knowing whether the suggested code fragments originate from open-source or proprietary repositories. This lack of provenance and traceability, in turn, can complicate audits, code reviews, and due diligence processes in professional development environments.

\begin{formal}\faPencil~
\textit{``The T\&C on GitHub do state that you grant them the right to `parse [your content] into a search index or otherwise analyze it on our servers'. It's not at all clear that this grants them the right to reproduce parts of your content (without credit) using Copilot. What about private repositories? ... It would be interesting to see if one can get Copilot to produce code that is in a private repo.'' (Hacker New user)}
\end{formal}

%Discussions on the potential legal and procedural risks of incorporating Copilot suggestions into proprietary or open source projects were also observed with high frequency. Developers often voiced uncertainty about the origin of Copilot-generated code and whether its reuse might introduce legal conflicts (e.g., copyright infringement or code plagiarism), especially within organizations that operate under strict audit and compliance policies. 

%As Copilot continues to be adopted in more structured and compliance-sensitive environments, the absence of transparent code provenance may hinder its acceptance or limit its integration into formal development pipelines. These findings underscore the need for platform-level design changes that improve traceability and reduce uncertainty for end users.

\subsubsection*{\textbf{\textcircled{\small{4}} Developer Trust Erosion and Overdependence on GenAI}}

Beyond code quality flaws and copyright issues, many developers raised concerns about their evolving relationship with GenAI coding tools, particularly the risk of placing excessive trust in Copilot’s outputs and gradually losing oversight of security-critical development tasks. These discussions reveal behavioral shifts in how developers perceive responsibility, with some expressing a sense of reduced accountability for the potential consequences of AI-generated code. Others noted a tendency to develop misplaced confidence in Copilot’s accuracy and reliability, especially when its suggestions appear syntactically correct or resemble familiar coding patterns.

%The discussions of this topic slightly overlap with those of Cluster 3 (``Legal, Licensing, and Attribution Ambiguity''). 

\begin{formal}\faPencil~
\textit{``In my opinion, Copilot is going to become one of those `perceived authorities' that have just enough legitimacy to be blindly trusted by the inexperienced, but not enough to actually be useful to the experts... The next generation of programmers will love the idea of Copilot. Instant gratification in the firm of a tool that can seemingly do your work for you. This will be dire consequences for their ability to code and think for themselves.'' (Hacker News user)}
\end{formal}

Some users have also warned of a potential decline in secure development habits due to Copilot's seemingly reliable and compelling suggestions. They emphasize that Copilot's ease of use and perceived correctness may lead developers to accept its outputs without sufficient scrutiny, particularly in fast-paced or low-awareness development settings. At the same time, many expressed growing mistrust toward such GenAI coding assistants, often shaped by prior experiences in which their recommendations introduced errors or insecure code in their projects.

\begin{formal}\faPencil~
\textit{``...(Copilot) is frankly dangerous for anything that is actually critical. I've found I can do it just as quickly, if not more so, because it's largely done right the first time instead of having to unpick Copilot's gibberish. People who have used Copilot are seemingly forever fixing bugs and having to deal with upset users, and so they banned it completely at work for all uses. The truth is that for serious work it is untrustworthy.'' (Reddit user)}
\end{formal}

%\textbf{Prompt Injection and Training Data Poisoning.}
%Lastly, the risks related to the manipulation of GitHub Copilot’s behavior via input prompts and the integrity of its training data were of concern to developers. This concern highlights vulnerabilities in the mechanisms that govern how Copilot learns and responds. 

%\textcolor{blue}{These findings underscore the need for stronger safeguards in both the training pipeline and runtime behavior of tools like Copilot. Addressing these concerns may require more \textbf{robust prompt sanitization}, detection of adversarial contributions to public repositories, and ongoing validation of model behavior across updates.}

\subsection{Nuances Across Platforms (RQ2)}\label{sec:nuances}
% \textit{RQ2: How do the security, privacy, and trust concerns vary across developer communities such as Reddit, Stack Overflow, and Hacker News?}\\
% \textcolor{blue}{Section 4.10}

%Our analysis of the 383 manually validated developers' post comments across the three platforms revealed the following high-level findings: (i) \textit{potential exposure of secrets and proprietary logic} was the most frequently discussed concern, highlighting ongoing unease with Copilot’s training integrity, (ii) \textit{insecure code suggestions} were a recurring issue irrespective of platform, (iii) \textit{prompt injection} and \textit{training data poisoning} were more specialized concerns mostly discussed by users focusing on long-term model safety, and (iv) \textit{developer trust} and \textit{behavioral erosion} were discussed on all platforms but were emotional and anecdotal on Reddit and procedural and architectural on Hacker News.

\begin{figure}[t]
    \centering
\includegraphics[width=\linewidth]{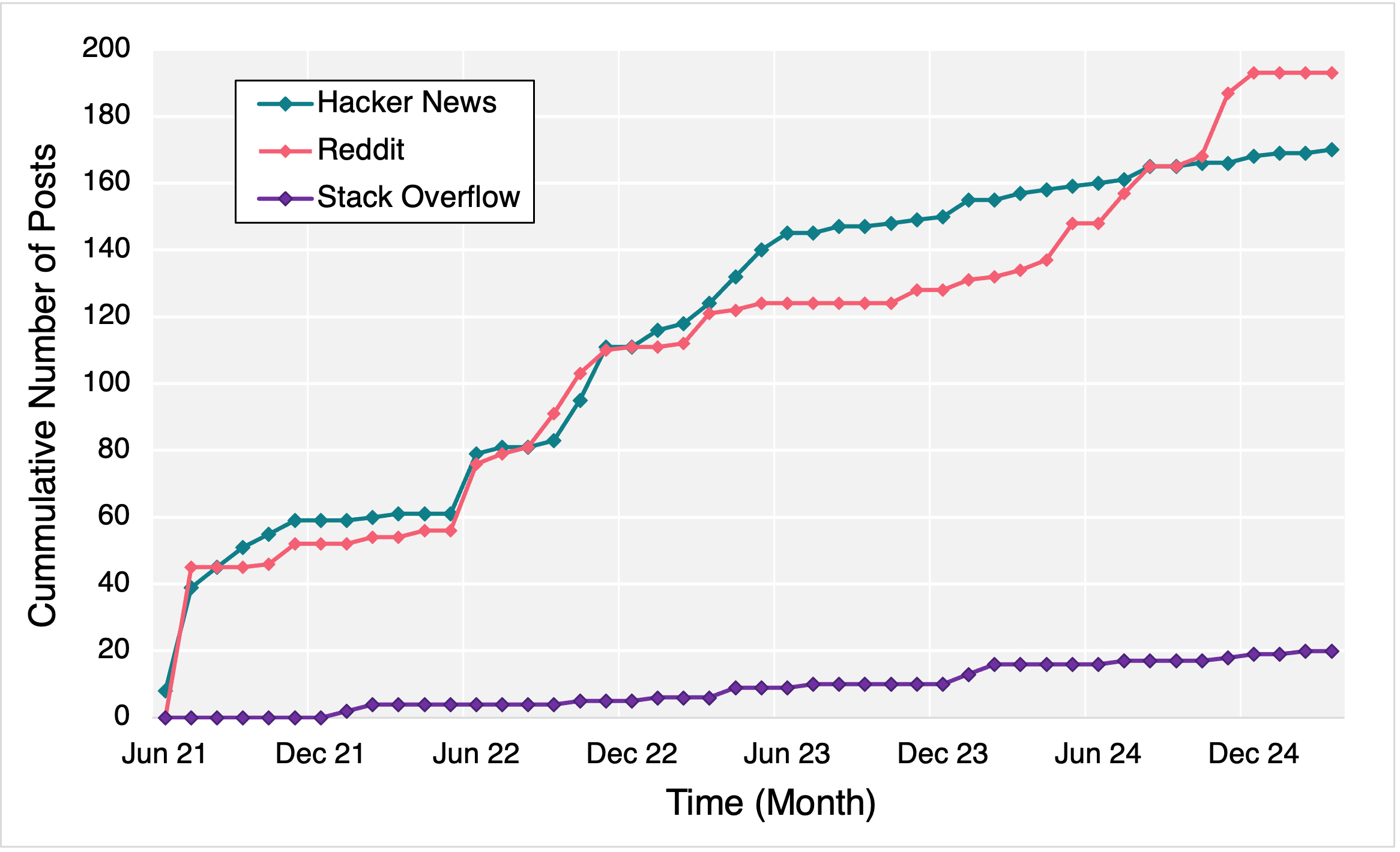}
    \caption{Cumulative number of relevant posts over time.}
    \label{fig:freq}
\end{figure}

Fig.~\ref{fig:freq} shows the cumulative frequency of security-related posts over time across Hacker News, Reddit, and Stack Overflow, revealing clear differences in engagement across platforms. While Stack Overflow exhibits a substantially lower volume of security-related discussions, this observation is consistent with prior work showing that security topics represent only a small fraction of overall Stack Overflow activity, which is otherwise dominated by general programming questions \cite{diaz2023cybersecurity}. A closer examination of each identified concern area revealed notable nuances in how developers discuss and approach these topics across the investigated platforms. As anticipated, the nature of discussions tends to reflect the culture and discourse norms of each community. On the one hand, conversations on Hacker News often reference external sources such as research articles, preprints, or news coverage. Stack Overflow threads are typically centered on concrete technical challenges and implementation issues, whereas Reddit discussions are broader and more conversational, frequently blending technical critique with personal opinions or ethical reflections. 

\begin{figure}[t]
    \centering
    \includegraphics[width=\linewidth]{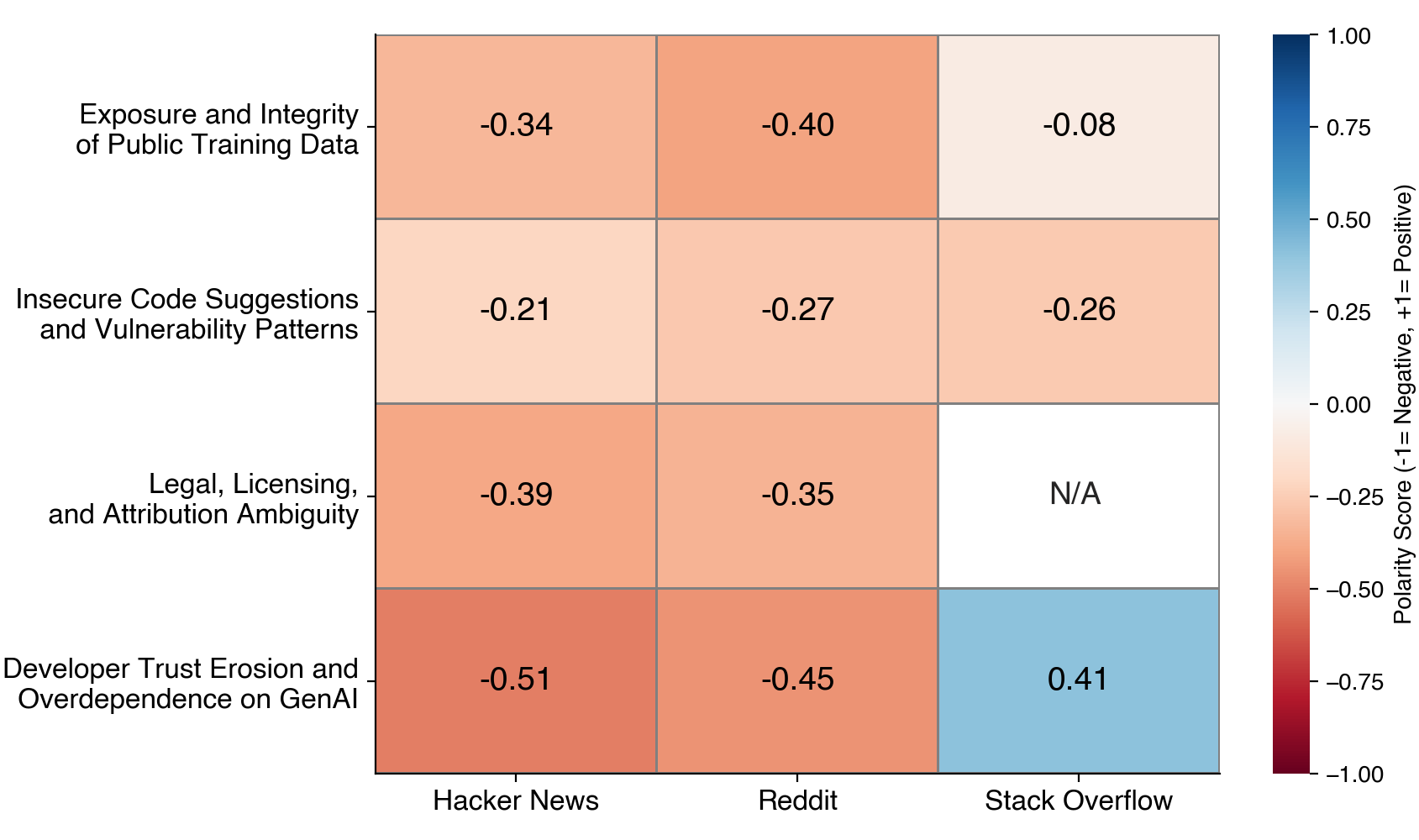}
    \caption{Sentiment Distribution Across Platforms.}
    \label{fig:sentiment}
\end{figure}

\subsubsection{Distribution of Concerns Across Platforms}\label{sec:topic_dist}

Concerns on \textit{``Exposure and Integrity of Public Training Data''} were more common on Reddit (41\%), followed by Hacker News  (45\%), but not so popular in Stack Overflow (14\%). Reddit users reported instances in which Copilot surfaced corporate identifiers, configuration tokens, or deprecated yet sensitive internal logic, whereas Hacker News posts contained more references to adversarial attacks (e.g., data poisoning and adversarial prompting). On Stack Overflow, the theme was often embedded in replies to security-related programming questions, such as \textit{``Why did Copilot suggest this GitHub token in a Django config?''} or \textit{``Is it safe to use these Copilot completions in production?''}.

References to \textit{``Insecure Code Suggestions and Vulnerability Patterns''} were the most prevalent overall, particularly on Reddit (38\%) and Hacker News (54\%), but less frequent on Stack Overflow (7\%). Reddit users often discussed how misplaced trust in Copilot can lead to copy-paste behavior among developers and, in turn, introduce security vulnerabilities (e.g., \textit{``It writes bad code with confidence. That's the scary part''}). On Hacker News, discussions focused on potential causes of Copilot's security limitations (e.g., low-quality training data) and on ways to mitigate potentially vulnerable suggestions (e.g., adding ad hoc security checks or limiting its use to repetitive, low-stakes tasks). Although we encountered relatively few Stack Overflow posts, these were short answers to specific code blocks suggested by Copilot that, for example, performed unsafe or unauthorized memory-access operations. Responses were often concise yet technically precise, typically providing corrected code samples and highlighting the root cause of the issue or safer implementation alternatives.

Concerns about \textit{Legal, Licensing, and Attribution Ambiguity} were predominant on Reddit (66\%) and Hacker News (34\%), but not on Stack Overflow. Discussions on both platforms addressed whether Copilot violated the GPL or MIT licenses, as well as GitHub's obligations to ensure compliance and transparency in this regard.  Anecdotes such as `\textit{`I copied something Copilot suggested and later realized it was verbatim from a popular repo''} were common, underscoring attribution ambiguity concerns. An alignment was also observed in the comments on \textit{``Developer Trust Erosion and Overdependence on GenAI''} extracted from Reddit (67\%) and Hacker News (30\%). Overall, users of both platforms expressed concerns about the long-term implications of relying on Copilot for routine programming tasks, questioning whether this could exacerbate knowledge gaps in software security if code generation becomes the default approach.

\subsubsection{Distribution of Sentiment Across Platforms} As shown in Fig.~\ref{fig:sentiment}, the sentiment expressed across all concern areas is generally skewed toward the negative end of the polarity scale. On Hacker News and Reddit, comments related to \textit{``Developer Trust Erosion and Overdependence on GenAI''} showed the strongest negative sentiment, suggesting critical attitudes among developers toward the prevalence of GenAI coding assistants and their implications for security practices. Similar patterns were observed for \textit{Legal, Licensing, and Attribution Ambiguity''} and \textit{Exposure and Integrity of Public Training Data''}, while \textit{Insecure Code Suggestions and Vulnerability Patterns''} displayed a milder yet still negative sentiment across all platforms. Finally, Stack Overflow posts generally scored near neutral on sentiment (except for one positive comment on \textit{``Developer Trust Erosion and Overdependence on GenAI''}), reflecting the platform's tendency to provide practical feedback and concrete code fixes rather than addressing broader legal, trust, or ethical~ aspects.

\section{DISCUSSION AND IMPLICATIONS} \label{sec:discussion}

To some extent, our findings align with prior investigations, suggesting that practitioners are well aware of the security limitations of GenAI-based coding assistants. At a general level, we observed that all emerging topics, except \textit{``Developer Trust Erosion and Overdependence on GenAI,''} exhibit negative sentiment across the scrutinized platforms. As mentioned in Section~\ref{sec:background}, such a tendency has also been observed previously in posts and discussions about ChatGPT in forums like Twitter \cite{koonchanok2024public} and Reddit \cite{ali2025understanding}. Nevertheless, while earlier studies have touched upon aspects related to insecure code suggestions and data leakage (e.g., \cite{klemmer2024using}), \textbf{two of the four concern areas identified in this work, namely \textit{``Legal, Licensing, and Attribution Ambiguity''} and \textit{``Developer Trust Erosion and Overdependence on GenAI''}, have not been thoroughly examined in prior work from a developer discourse perspective}. Furthermore, even for areas previously surfaced in earlier investigations (i.e., \textit{Exposure and Integrity of Public Training Data''} and \textit{``Insecure Code Suggestions and Vulnerability Patterns''}), our study provides insights into their practical relevance through the frequency and technical depth with which they are discussed across online forums.

\subsubsection*{\textbf{Trust and Training Data Governance}} Overall, the identified concern areas highlight critical trust gaps in the design of GenAI coding assistants and call for greater transparency, stronger accountability mechanisms, and stronger security safeguards. On the one hand, developers' discussions around \textit{``Exposure and Integrity of Public Training Data''} highlight systemic failures in the way technologies like GitHub Copilot handle sensitive data, surfacing risks (e.g., unintended memorization, data leakage, and the potential misuse of publicly available repositories) which may compromise both security and intellectual property. Without stricter transparency, sanitization of training data, and real-time safeguards such as secret detection or prompt-based warnings, trust in Copilot's deployment within security-critical environments is likely to remain constrained \cite{wu2025empirical}. Novel LLM applications of \textit{Machine Unlearning} techniques \cite{liu2025rethinking} could help mitigate some of these issues by enabling models to selectively remove the influence of sensitive data types while retaining their overall knowledge generation capabilities. Moreover, concerns about data poisoning and prompt injection attacks underscore the need for stronger safeguards in both the training pipeline and the runtime behavior of these tools. Addressing these concerns may require more robust system prompt sanitization, detection of adversarial contributions to public repositories, and ongoing validation of model behavior across updates~\cite{geroimenko2025key}.

\subsubsection*{\textbf{Developer-Centered Security Awareness}} As described in Section~\ref{sec:background}, issues around \textit{``Insecure Code Suggestions and Vulnerability Patterns''} have attracted significant research attention in recent years. Corrective approaches, such as secure prompt optimization \cite{nazzal2024promsec}, aim to reactively counteract the generation of insecure code by refining LLM inputs \cite{xu2024large}. Still, these methods are more likely to be adopted by security-savvy practitioners, as they are not yet embedded by design in the architecture or underlying models of GenAI coding assistants. Fine-tuning such models for secure code generation could help bridge this gap; however, recent investigations indicate that, although promising, it cannot fully replace developers' judgment in its current form \cite{li2024fine}. As reflected in the discussions around \textit{``Insecure Code Suggestions and Vulnerability Patterns''}, strengthening developers' critical awareness of quality and security issues in LLM-generated code remains essential. Addressing these concerns may require tool-level interventions, digital nudges (e.g., quality cues) \cite{serafini2025exploring}, or embedded reminders that encourage developers to maintain a reflective and security-conscious assessment of the code produced by GenAI assistants.

%for actions to safeguard both end users' ... and preserve developer's autonomy ...

%\textcolor{blue}{ON TOPIC 2: This theme underscores a central trade-off in Copilot’s design and usage. Although it significantly accelerates code generation and scaffolding, it often \textbf{omits essential security features} and defaults to implementations that lack \textbf{defensive programming practices}. The issue lies not only in what Copilot suggests, but also in what it fails to specifically suggest the security measures that should accompany core functionality. Because Copilot reflects common patterns in its training data, which are not always secure or up to date, it risks perpetuating unsafe practices, particularly among less experienced developers. These observations raise critical questions about the adequacy of Copilot's suggestions in real-world, security-sensitive software development environments.}

\begin{figure}[b]
\vspace{-2ex}
\begin{tcolorbox}[width=\linewidth,colback={gray!20},boxsep=1pt]    
\textbf{KEY IMPLICATIONS AND TAKEAWAYS.}
\begin{enumerate}[leftmargin=14pt]
\item Salient mismatches between developers’ expectations and the current technical affordances of GenAI coding assistants emerged across four security-critical areas related to \textit{code licensing}, \textit{training data leakage}, \textit{insecure code implementations}, and \textit{developers’ autonomy}.\vspace{1ex}
\item GenAI coding assistants should promote a savvy behavior among developers by supporting a \textit{security-conscious assessment} of the recommended code (e.g., through code quality cues and behavioral interventions).
\vspace{1ex}
\item Clear documentation of \textit{licensing provenance} should accompany GenAI-generated code to help developers and organizations assess reuse and compliance risks.\vspace{1ex}
\item Transparency and traceability in GenAI-based code recommendations should be strengthened, alongside safeguards against \textit{data poisoning attacks} and the \textit{memorization of sensitive data} during training.\vspace{1ex}
\item Altogether, the identified concern areas call for coordinated and adaptive safeguards spanning \textit{model design}, \textit{tooling}, and \textit{developer interaction}, rather than isolated, one-size-fits-all solutions.
%\item Future work should seek to enhance the transparency and traceability of GenAI-based code recommendations, while preventing \textit{data poisoning attacks} and the \textit{memorization of sensitive data} during training.
\end{enumerate}
\end{tcolorbox}%\vspace{-3ex}
\end{figure}
%\item The diversity of concerns observed across platforms suggests that security challenges are context-dependent , underscoring the need for adaptive safeguards rather than one-size-fits-all solutions.

\subsubsection*{\textbf{Provenance, Transparency, and Traceability}} Finally, \textit{``Legal, Licensing, and Attribution Ambiguity''} concerns underscore the need for model-level design changes in GenAI coding assistants to improve traceability and reduce uncertainty for end users. Without appropriate mechanisms for attribution, license tagging, or traceable output, practitioners are left with limited instruments to assess the legal and ethical implications of the proposed code. As Copilot continues to be adopted across corporate and compliance-sensitive environments, the absence of transparent code provenance may hinder its acceptance or limit its integration into real-world development pipelines. Recent work has begun to examine license compliance in LLM-generated code from a technical evaluation perspective. For instance, Xu et al. proposed LiCoEval ~\cite{xu2025licoeval}, a benchmark to assess LLMs' ability to provide accurate license information for generated outputs under \textit{striking similarity} conditions with open-source code. While such approaches provide important foundations for detecting provenance risks, they are still in their infancy and require broader empirical validation at scale that considers how practitioners in real-world development settings perceive, interpret, and act on licensing uncertainties.

%\item Four major concern areas were identified across the analyzed platforms, covering issues related to \textit{code licensing}, \textit{training data leakage}, \textit{insecure code implementations}, and \textit{developers' overreliance} in GenAI code suggestions. 

%\item Hacker News and Stack Overflow posts mostly focused on insecure code suggestions with some of the former also mentioning adversarial attacks. Reddit discussions addressed problems related to code licensing more often. 

\section{THREATS TO VALIDITY} \label{sec:limitations}
%Despite our best efforts, our study has limitations that could impact our findings. Following Runeson et al.~\cite{runeson2009guidelines} classification, we discuss the potential risks to this study and the relevant measures we took to mitigate them.

\textbf{Construct Validity.} The analyzed discussions may not fully capture the security concerns of the broader developer population. Instead, they may reflect the perspectives of developers active on these platforms, while omitting viewpoints that might emerge in offline or less publicly accessible settings. We sought to mitigate this threat by integrating evidence from multiple sources, thereby enhancing the diversity of topics discussed and the variation in user post styles. On the other hand, the keyword-based approach used to identify relevant online discussions may have overlooked cases expressed using alternative terminology. While the breadth of the keyword set provides reasonable coverage of security-related posts, the selection criteria applied to identify suitable threads and subreddits may have introduced biases. In particular, relying on popularity metrics (e.g., number of comments or user activity) ensured relevance but may have favored dominant viewpoints over more nuanced or less visible perspectives.

\textbf{Internal Validity.} Concerns around GitHub Copilot may not fully generalize to other GenAI coding assistants, although its popularity suggests broader relevance. Furthermore, while BERTopic provides cohesive discussion clusters, their interpretation involved a manual and subjective process based on representative samples. The identification of security-related discussions was also primarily conducted by a single researcher, which may introduce bias. To mitigate this, a second author with expertise in software security validated the dataset and the derived concern areas.

\textbf{External Validity.} 
As mentioned earlier, our findings are based on a relatively small sample spanning three platforms and centered on security-related discussions of GitHub Copilot. Such a sample captures a valuable yet narrow share of developers' concerns voiced online. In turn, we cannot generalize our results to the entire community of software practitioners and users of GenAI coding assistant technologies. In fact, it can only account for the experiences of those who are, coincidentally, active in Reddit, Stack Overflow, or Hacker News. In turn, the insights we gained should be seen as preliminary and motivate further research in this area. Moreover, future investigations should aim for larger and gender-diverse samples to capture the perspectives of underrepresented groups (e.g., women and gender-diverse individuals).

\textbf{Conclusion Validity.} Finally, the sentiment and thematic nuances observed across the three analyzed platforms (Section~\ref{sec:nuances}) only remain valid on a descriptive level, as our results did not yield insights into their statistical significance. The observed differences may also be influenced by the choice and configuration of algorithms used for clustering and sentiment analysis. We configured these algorithms in accordance with established best practices to ensure the relevance and reproducibility of our findings.

\section{CONCLUSION AND FUTURE WORK} \label{sec:conclusion}

%In this work, we have identified and characterized some key security concern areas related to the use of GenAI for software development. 
%Our findings, derived from the analysis of public discussions on GitHub Copilot across three popular platforms, indicate significant security gaps in the current design of GenAI coding assistants. 

Our findings, derived from an analysis of public discussions on GitHub Copilot across three popular platforms, highlight recurring security-related challenges and areas of concern associated with the current design of GenAI coding assistants. In particular, issues related to legal compliance, data governance, and the ability to assess and mitigate potential vulnerabilities in code suggestions appear to be at the forefront of developers' security concerns. While technical issues, such as the potential exposure of sensitive training data and the generation of vulnerable code, feature prominently in these discussions, there is also a strong sense of uncertainty regarding attribution, licensing, and intellectual property. Furthermore, discussions of overreliance on GenAI suggestions and the gradual erosion of developers' security skills reveal underlying tensions among automation, human judgment, and responsibility in software development.

Overall, our results suggest that while GenAI offers substantial productivity benefits for software practitioners, its deployment in security-sensitive contexts remains contentious. The identified concern areas, which often echoed technical, legal, and user-centered challenges previously reported in the literature, call for concrete design interventions that support developers’ security-critical decisions in a transparent and compliant manner. Our findings indicate that incorporating strong default security safeguards into the foundations of LLM coding agents could contribute substantially in this regard. Moreover, providing detailed quality and licensing feedback alongside code recommendations could further promote security-savvy decision-making within GenAI-augmented software processes. Still, dedicated research efforts are needed to translate these prospective directions into practical, empirically validated design guidelines.

Several directions for future work emerge from the results of this study. One relates to the collection of additional evidence from other platforms, such as GitHub Community Discussions, to better characterize and expand the identified areas of concern. In line with this, further research should consider other coding assistants, such as Claude.AI and Google's Gemini Code Assist, to identify nuances across multiple solutions. Surveys or semi-structured interviews with security experts could help address some concern sub-areas with greater detail, particularly those, such as adversarial attacks, that are supported by limited evidence in our dataset. Finally, the results of this work could aid the design of behavioral interventions or nudges to promote security-savvy decisions in GenAI-augmented coding environments. Particularly, to provide practitioners with timely and actionable feedback on the quality and provenance of code suggestions.

%Extend beyond copilot
%Perform longitudinal studies
%Shape interventions and test them in a controled experimental setting.

%Still, to the extent of our knowledge, little evidence has been gathered around the security of GenAI coding assistants from a developer-centered perspective, which calls for further empirical investigations ...

%However, "content and form of output is less predictable than conventional ML due to generative nature, opaqueness of models makes it difficult to specify adequate criteria for validation...Lack of certifications and quality standards" \cite{dolata2024development} 

%An exploratory case study by YYY explored the interactions of GenAI within a software development team and revealed the need for validation in security-critical tasks.

%Recent investigations have began to explore developers' perceptions of GenAI technologies 

%Recent investigations have began to explore how software developers themselves engage with GenAI technologies, particularly in relation to security risks in their day-to-day coding activities.

%studies using Q&A platforms and more centred in SW dev.
% 2024_nguyen: focus groups (done)
% 2025_banh: interviews with developers
% 2024_wu: Interviews (case study) with developers (done)
% 2024_dolata: study with freelancers (some of them voiced some challenges in security)
% 2025_zegers: Reddit study. Topic modeling

% All in all, there is no study dedicated to capture the privacy and security perspectives of developers with regard to GenAI-based coding assitants (specific to this type of GenAI application).

\section*{ACKNOWLEDGMENTS}

This work was partly supported by the European Union under grant No. 101120393 (Sec4AI4Sec).

\bibliographystyle{ACM-Reference-Format}
\bibliography{references}

\end{document}